\documentstyle[preprint,aps,epsf]{revtex}
\tighten

\begin{document}

\draft

\title{
Search for Three-Nucleon Force Effects in Two-Body Photodisintegration
of $\bbox{^3}$He ($\bbox{^3}$H) and in the Time
Reversed Proton-Deuteron Radiative Capture Process
}

\author{
R.~Skibi\'nski$^1$,
J.~Golak$^{1,2}$,
H.~Kamada$^3$
H.~Wita\l{}a$^1$,
W.~Gl\"ockle$^2$,
A.~Nogga$^4$.
}
\address{$^1$M. Smoluchowski Institute of Physics, Jagiellonian University,
                    PL-30059 Krak\'ow, Poland}
\address{$^2$Institut f\"ur Theoretische Physik II,
         Ruhr Universit\"at Bochum, D-44780 Bochum, Germany}
\address{$^3$ Department of Physics, Faculty of Engineering,
   Kyushu Institute of Technology,
   1-1 Sensuicho, Tobata, Kitakyushu 804-8550, Japan}
\address{$^4$ Department of Physics, University of Arizona, Tucson,
              Arizona 85721, USA}

\date{\today}
\maketitle

\begin{abstract}
Faddeev calculations have been performed for nucleon-deuteron
photodisintegration
of $^3$He ($^3$H) and proton-deuteron radiative capture. 
The bulk of the results is based on the AV18 
nucleon-nucleon force
and the Urbana IX three-nucleon force together with explicit
exchange currents or applying the Siegert approach. Three- nucleon force
effects are predicted for both processes and are qualitatively supported
by available data.
\end{abstract}

\narrowtext

\section{Introduction}
\label{secIN}

Three-nucleon (3N) forces come more and more into the focus of few-nucleon
studies. Pure 3N continuum measurements at the accelerator facilities
IUCF~\cite{ref.IUCF}, KVI~\cite{ref.KVI}, RIKEN~\cite{ref.RIKEN}
and RCNP~\cite{ref.RCNP} are performed  around
100-200 MeV nucleon laboratory energies with the aim to confront
data to theoretical  predictions   based on modern high-precision
nucleon-nucleon (NN) forces only~\cite{ref.Report}.
Clear-cut discrepancies
 for certain 3N observables
against all those predictions
can be considered to be
good candidates for 3N force (3NF) effects. Thereby
 the  theoretical investigations are  based on
 numerically precise solutions of the 3N Faddeev equations.
Then adding  present day  3NF models
 and comparing to those data   one tries  to explore their strength
and spin structure~\cite{ref.spinstructure}. Right now
these 3NF models are the 2$\pi$-exchange
Tucson Melbourne (TM)~\cite{ref.TM}, a modified
version thereof, TM'~\cite{ref.modTM}, which is  closer to chiral symmetry,
and the Urbana~IX~\cite{ref.urbanaIX}
forces.
 Another path to learn about 3NF's is the study of the low lying
spectra of light nuclei as performed in Greens function Monte Carlo
calculations~\cite{ref.GFMC1}.
 The inclusion of 3$\pi$-exchange ring diagrams with intermediate
$\Delta$'s on top of the Urbana IX 3NF appears to be rather promising to
improve the theoretical description of the spectra~\cite{ref.spectra}.
In all those investigations there is clear evidence
found that present day NN forces alone fail to describe many of the studied observables
and adding the presently available three- nucleon force models moves theory in
 the right direction.

A recent approach towards nuclear dynamics is based on chiral perturbation theory, which
is closely linked to QCD and develops nuclear forces in a systematic and controlled manner~\cite{wei1}.
In that scheme which treats multi- pion exchanges explicitly and incorporates short
range processes in the form of contact forces of increasing  chiral dimensions also 3N forces
are predicted and this consistently to NN forces. In~\cite{epe1} it has been demonstrated that like
in the conventional approaches mentioned above 3N forces are unavoidable to predict
binding energies of three- and four- nucleon nuclei as well as to remove discrepancies in
certain 3N scattering observables.

Electromagnetically induced reactions in the 3N system should also show
effects of 3NF's. 
Since both 3N bound and scattering states enter into the nuclear matrix elements for
photon induced processes and both types of states are affected by 3N forces it would be
surprising if the various response functions for these reactions would be unaffected. In
principle just by the continuity equation 3N forces lead also unavoidably to 3N
currents. It is a quantitative question based on current choices of nuclear force models to
reveal signatures by switching on and off 3N forces. If certain observables are linked to
binding energies and if all modern NN forces including the most recent ones based on chiral
perturbation theory are unable to predict the experimental bound state energies but the
inclusion of 3N forces is, we leave it to the reader to decide whether the changes in
those observables are called 3N force effects or just binding effects. Apparently
under these circumstances both are  tightly bound together. It is only with oversimplified
 toy model
NN forces  which do not describe the rich  NN data set that possibly experimental bound
state energies can be achieved. Conclusions based on those models should be taken with
caution.
The search for three- nucleon force effects in electromagnetically induced processes has
been started before. For recent references see~\cite{car1,car2,sch1,efr1}.
It is the aim of this paper to
investigate
the nucleon-deuteron (Nd)
photodisintegration of $^3$He and $^3$H
as well as the time reversed
proton-deuteron (pd) capture process
using modern NN forces and various 3NF
models.

The single nucleon current operator is supplemented by exchange
currents either in the form of  the Siegert  approximation  or by
 explicitly including MEC's of the $\pi$- and $\rho$-like nature.
 The treatment is carried through nonrelativistically, though presumably
 some of the data that we analyze, require at least relativistic
 corrections.

Two-body photodisintegration of $^3$He ($^3$H) has a long history. Barbour
and Phillips~\cite{ref.Barbour&Phillips} found that the incorporation of the
interacting 3N continuum is crucial for the understanding of that process.
They solved the 3N Faddeev equations, at that time of course based on simple
finite rank forces.
This was taken up again more consistently by Gibson and
Lehman~\cite{ref.Gibson&Lehman}
treating the 3N bound state and the final 3N  continuum on equal footing.
More recently the Bonn group~\cite{ref.Bonn,ref.Bonn.2} used more modern NN
forces represented in
finite rank form. They analyzed quite a few data and
pointed out a correlation of a certain cross
section peak height with the triton binding energy, an issue which we
shall also address but now in the context of 3NF effects.
All the work mentioned  relied on the  Siegert approximation.
The current was restricted to the dominant $E_1$ multipole~\cite{ref.Bonn}
or to the $E_1$ and $E_2$ multipoles~\cite{ref.Bonn.2}.
In a recent paper~\cite{Trento} a benchmark was set on the total
3N photodisintegration cross section.
There two quite different approaches, the Faddeev one and a hyperspherical harmonic expansion
together with the Lorentz integral transform method were compared  to each other
using AV18 together with
Urbana IX and reached a very good agreement. This documents the technical maturity of
advanced present day approaches.

Also for the pd capture process many experimental and theoretical studies
have been performed in the past. We refer to
\cite{ref.Bonn.2,ref.14,ref.15} for references.
Specifically we want to point to the theoretical investigations by
Schadow {\em et al.}~\cite{ref.Bonn.2},
Fonseca and Lehman~\cite{ref.Fonseca&Lehman} and to recent studies
at very low energies by Viviani {\em et al.}~\cite{ref.Schiavilla.etal}.

The present investigation is restricted to  nucleon-  deuteron fragmentations
 in relation to $^3$He ($^3$H) and
we refer to a forthcoming study for 3N fragmentations.

The paper is organized as follows. In Sec.~II we briefly review our
formalism and the dynamical ingredients. Our results for Nd
photodisintegration are presented in Sec.~III together with available
data. The pd capture observables are  discussed  in Sec.~IV.
We summarize in Sec.~V.

\section{Formalism}
\label{secII}

We evaluated photodisintegration and pd capture before
in~\cite{ref.15} and~\cite{ref.16,ref.17},
always using Faddeev-like integral equations. In~\cite{ref.17} we formulated pd
capture based on NN forces alone. There the Faddeev-like integral
equation is identical to the one for Nd scattering. This is because
in pd capture
the 3N scattering  state $\mid \Psi^{(+)} \rangle$ enters directly.
On the other hand
in $^3$He photodisintegration the 3N scattering state $\langle \Psi^{(-)} \mid$
is involved like in electrodisintegration of $^3$He. The way to derive
Faddeev-like integral equations in the latter cases is to apply
the adjoint Moeller wave  operator
entering the nuclear matrix element to the right, namely onto the
electromagnetic current operator and the $^3$He bound state~\cite{ref.18}. This has the
very big advantage that the driving term of that Faddeev-type integral
equation is fully connected, namely proportional to the $^3$He bound state.
Because of the formal identity of the nuclear matrix elements for
photodisintegration and electron induced processes the same
Faddeev-like integral equation is applicable. In the two cases
only the components of the current operator in the driving term
 have to be chosen
appropriately. Now using time reversal symmetry one can relate
the matrix elements for pd capture and photodisintegration, which are
evaluated quite differently. This is a highly
nontrivial numerical test for the various complex numerical steps
involved.

In~\cite{ref.15} we added a 3NF in the evaluation of the pd capture process and
applied it to cross sections and several spin observables. The
formalism was straightforward since we could use directly the
Faddeev-like integral equation for Nd scattering
including 3N forces as derived in~\cite{Newmethod}.
In the same paper~\cite{ref.15} we compared Siegert approximation to the explicit use of a
restricted but possibly dominant set of mesonic exchange currents.

Now for two- and three-body photodisintegration we would also like to formulate
 an extension including 3N forces.
We shall proceed as follows. The nuclear matrix element for
Nd-photodisintegration of a 3N bound state has the following form

\begin{eqnarray}
N_\tau^{\rm Nd} \equiv
\langle \Psi_{\vec q}^{(-)} \mid j_\tau (\vec Q ) \mid \Psi_{\rm bound} \rangle ,
\label{eq:N}
\end{eqnarray}
where $\vec q$ is the asymptotic relative momentum between the proton and
the deuteron, and $ j_\tau (\vec Q )$ is the component of the
3N current operator.
The scattering state $\langle \Psi_{\vec q}^{(-)} \mid $   can be Faddeev decomposed

\begin{eqnarray}
 \langle \Psi_{\vec q}^{(-)} \mid  =
 \langle \psi_{\vec q}^{(-)} \mid ( 1 + P ) ,
\label{eq:decomp}
\end{eqnarray}
where $P$,  according to our standard notation~\cite{gloecklebook}, is the sum of a cyclical
and anticyclical permutation. The Faddeev amplitude  $  \langle \psi_{\vec q}^{(-)} \mid $
obeys the Faddeev equation~\cite{Newmethod}

\begin{eqnarray}
 \langle \psi_{\vec q}^{(-)} \mid  =
 \langle \phi_{\vec q} \mid  +
 \langle \psi_{\vec q}^{(-)} \mid \left( P t G_0  + ( 1 + P ) V_4^{(1)} G_0 ( t G_0 +1) \right) .
\label{eq:fad}
\end{eqnarray}
Here the channel state $ \langle \phi_{\vec q} \mid $ enters, which is
a product of a deuteron wave function and a momentum
eigenstate of the remaining nucleon,
the NN t-operator $t$ acting on nucleons 2 and 3,
the free 3N propagator $G_0$,
the permutation operator $P$
and $ V_4^{(1)} $, the part of a 3N force, which singles out particle 1.
For our notation see~\cite{gloecklebook} and \cite{ref.Report}.

Using Eqs.~(\ref{eq:N}), (\ref{eq:decomp})
and (\ref{eq:fad}) we can write the nuclear matrix
element as

\begin{eqnarray}
N_\tau^{\rm Nd} =
\langle \phi_{\vec q} \mid  ( 1 - K )^{-1} ( 1 + P ) j_\tau (\vec Q )
\mid \Psi_{\rm bound} \rangle ,
\label{eq:N2}
\end{eqnarray}
where $K$ is the kernel of the integral equation~(\ref{eq:fad}). We introduce

\begin{eqnarray}
\mid U \rangle  \equiv ( 1 - K )^{-1} ( 1 + P ) j_\tau (\vec Q ) \mid \Psi_{\rm bound} \rangle
\label{eq:U}
\end{eqnarray}
or explicitly the integral equation
\begin{eqnarray}
\mid U \rangle  = ( 1 + P ) j_\tau (\vec Q ) \mid \Psi_{\rm bound} \rangle +
\left( P t G_0  + ( 1 + P ) V_4^{(1)} G_0 ( t G_0 +1) \right) \mid U \rangle .
\label{eq:U2}
\end{eqnarray}
This form is not yet suitable for numerical applications because of the
presence of $P$ to the very left. This has already been noted at the
very beginning of our numerical 3N studies using nuclear forces without
finite rank representations~\cite{Boemelburg}. To rewrite Eq.~(\ref{eq:U2})
 into a suitable form
we use the following obvious identities

\begin{eqnarray}
 ( 1 + P ) = \frac12 P ( 1 + P )
\label{eq:P1}
\end{eqnarray}
\begin{eqnarray}
\frac12 P ( P - 1 ) = 1
\label{eq:P2}
\end{eqnarray}
and obtain

\begin{eqnarray}
 ( P - 1 ) \mid U \rangle  =
( P - 1 ) ( 1 + P ) j_\tau (\vec Q ) \mid \Psi_{\rm bound} \rangle + \cr
( P - 1 ) P  \left( t G_0  + \frac12 ( 1 + P ) V_4^{(1)} G_0 ( t G_0 +1) \right)
\frac12 P ( P - 1 ) \mid U \rangle .
\label{eq:PU2}
\end{eqnarray}
This  Faddeev-like integral equation is suitable for numerical implementations
and has the form

\begin{eqnarray}
\mid \tilde{U} \rangle  =
( 1 + P ) j_\tau (\vec Q ) \mid \Psi_{\rm bound} \rangle + \cr
\left( t G_0 P + \frac12 ( 1 + P ) V_4^{(1)} G_0 ( t G_0 +1) P \right)
\mid \tilde{U} \rangle
\label{eq:Utilde}
\end{eqnarray}
with

\begin{eqnarray}
\mid \tilde{U} \rangle  \equiv  ( P - 1 ) \mid U \rangle .
\label{eq:UtildeU}
\end{eqnarray}
Then the nuclear matrix element results as
\begin{eqnarray}
N_\tau^{\rm Nd} =
\frac12 \langle \phi_{\vec q} \mid  P \mid \tilde{U} \rangle .
\label{eq:N3}
\end{eqnarray}
In view of a forthcoming paper we also describe now  the treatment of the
complete 3N break-up process. The nuclear matrix element is

\begin{eqnarray}
N_\tau^{\rm 3N} \equiv \langle \Psi_{\vec p \, \vec q}^{(-)}
\mid j_\tau (\vec Q ) \mid \Psi_{\rm bound} \rangle .
\label{eq:N3N}
\end{eqnarray}
The asymptotic momenta of the three nucleons are given by standard
Jacobi momenta $\vec p$  and $\vec q$~\cite{gloecklebook}.
The Faddeev amplitude corresponding to the scattering state in Eq.~(\ref{eq:N3N})
is now defined via

\begin{eqnarray}
 \langle \psi_{\vec p \, \vec q}^{(-)} \mid \  =  \
^{(-)}\langle \vec p \, \vec q \mid \ +  \ \langle \psi_{\vec p \, \vec q}^{(-)} \mid K ,
\label{eq:fad3N}
\end{eqnarray}
where $K$ is the same kernel as used before in Eq.~(3) and

\begin{eqnarray}
^{(-)}\langle \vec p \, \vec q \mid \ \equiv \
 \langle \phi_0 \mid ( t G_0 + 1 ) .
\label{eq:pqmin}
\end{eqnarray}
It is to be noted that the free two-body subsystem state in $  \langle \phi_0 \mid $
is properly antisymmetrized. Here  $  \langle \phi_0 \mid $ is the free 3N
state. Following the same  steps as above one
ends up with

\begin{eqnarray}
N_\tau^{\rm 3N} =
\frac12  \langle \phi_0 \mid ( t G_0 + 1 ) P \mid \tilde{U} \rangle ,
\label{eq:N3NN}
\end{eqnarray}
where $ \mid \tilde{U} \rangle $ is as given above.

In the actual numerical calculation we used, however, another form, which
for the purpose of completeness,  we would also like to present here. The
reason for that is that at the time of the installation,   the very heavy
numerical tasks were  more easily performed with already existing
 building blocks.
That alternative forms are for the Nd-break up

\begin{eqnarray}
N_\tau^{\rm Nd} =
\langle \phi_{\vec q} \mid  ( 1 + P ) \mid
j_\tau (\vec Q ) \mid \Psi_{\rm bound} \rangle +
\langle \phi_{\vec q} \mid  P \mid U' \rangle
\label{eq:Nnew}
\end{eqnarray}
and for the 3N break up

\begin{eqnarray}
N_\tau^{\rm 3N} =
\langle \phi_0 \mid ( 1 + P )
j_\tau (\vec Q ) \mid \Psi_{\rm bound} \rangle +
\langle \phi_0 \mid t G_0 ( 1 + P )
j_\tau (\vec Q ) \mid \Psi_{\rm bound} \rangle + \cr
\langle \phi_0 \mid P \mid U' \rangle +
\langle \phi_0 \mid t G_0 P \mid U' \rangle .
\label{eq:N3new}
\end{eqnarray}
The Faddeev-like integral equation  for $ \mid U' \rangle $ reads then

\begin{eqnarray}
\mid U' \rangle  =
\left( t G_0 + \frac12 ( 1 + P ) V_4^{(1)} G_0 ( t G_0 +1) \right)
( 1 + P ) j_\tau (\vec Q ) \mid \Psi_{\rm bound} \rangle + \cr
\left( t G_0 P + \frac12 ( 1 + P ) V_4^{(1)} G_0 ( t G_0 +1) P \right)
\mid U' \rangle .
\label{eq:Uprime}
\end{eqnarray}

The equivalence between the matrix elements  (\ref{eq:N3}) and (\ref{eq:N3NN})
on the one hand and (\ref{eq:Nnew}) and (\ref{eq:N3new}) on the other hand
is demonstrated as
follows. From Eqs.~(\ref{eq:Utilde}) and (\ref{eq:Uprime}) we have

\begin{eqnarray}
\mid \tilde{U} \rangle  - \mid U' \rangle  =  \cr
(1 - K )^{-1}
\left(
( 1 + P ) j_\tau (\vec Q ) \mid \Psi_{\rm bound} \rangle  -
t G_0 ( 1 + P ) j_\tau (\vec Q ) \mid \Psi_{\rm bound} \rangle  - \right. \cr
\left. \frac12 ( 1+ P ) V_4^{(1)} G_0 ( t G_0 +1) ( 1 + P ) j_\tau (\vec Q )
\mid \Psi_{\rm bound} \rangle \right) = \cr
(1 - K )^{-1} \left( 1 - t G_0 - \frac12 ( 1 + P ) V_4^{(1)} G_0 ( t G_0 +1) \right)
( 1 + P ) j_\tau (\vec Q ) \mid \Psi_{\rm bound} \rangle .
\label{eq:Udiff}
\end{eqnarray}
Using again Eq.~(\ref{eq:P1}) and the form of the kernel $K$ this is

\begin{eqnarray}
\mid \tilde{U} \rangle  - \mid U' \rangle  =  \cr
(1 - K )^{-1} \frac12
( P -1 + 1 - K ) ( 1 + P ) j_\tau (\vec Q ) \mid \Psi_{{\rm bound}} \rangle  = \cr
\frac12 (1 - K )^{-1} (P -1) ( 1 + P ) j_\tau (\vec Q ) \mid \Psi_{\rm bound} \rangle +
\frac12 ( 1 + P ) j_\tau (\vec Q ) \mid \Psi_{\rm bound} \rangle = \cr
\frac12 (1 - K )^{-1} ( 1 + P) j_\tau (\vec Q ) \mid \Psi_{\rm bound} \rangle +
\frac12 ( 1 + P) j_\tau (\vec Q ) \mid \Psi_{\rm bound} \rangle \equiv \cr
\frac12 \mid \tilde{U} \rangle  + \frac12 ( 1 + P) j_\tau (\vec Q ) \mid \Psi_{\rm bound} \rangle .
\label{eq:Udiff2}
\end{eqnarray}
Thus

\begin{eqnarray}
\frac12 \mid \tilde{U} \rangle  - \mid U' \rangle  =
\frac12 ( 1 + P) j_\tau (\vec Q ) \mid \Psi_{\rm bound} \rangle .
\label{eq:Udiff3}
\end{eqnarray}
Now it is a simple task to verify that the two expressions for the 3N
break-up amplitude (16) and (18) are identical.
This is also true for the two Nd
break-up amplitudes (12) and (17).
It is efficient to evaluate also the pd capture process using time reversal
in terms of the formalism just described. This is what we do in this paper.

\section{Results for pd (nd) Photodisintegration of
$\bbox{^3}$He ($\bbox{^3}$H).}
\label{secIII}

We use the AV18 NN force~\cite{ref.AV18}
combined with the Urbana~IX 3N force~\cite{ref.urbanaIX}.
By construction that force combination describes the $^3$H binding energy
correctly. It overbinds however slightly the $^3$He bound state energy by
21 keV~\cite{Noggaphd}.
For the convenience of the reader we cite the theoretical
binding energies : -7.628 MeV  for $^3$H and AV18 alone, -8.48 by construction including Urbana IX,
-6.917 MeV for $^3$He and AV18 together with the Coulomb force and -7.739 MeV including
 in addition Urbana IX. 
The latter value is to be compared to the experimental value -7.718 MeV which is slightly
different, but this should be of no significance for the present studies. Unfortunately we
are still unable to include the Coulomb force into the pd continuum, which causes an
inconsistency of unknown magnitude. Of course at least for the higher energies studied in
this paper we expect minor Coulomb force effects as it is supported by pure pd scattering
investigations~\cite{ref.Report}.
Our calculations are fully converged by choosing total two-
nucleon angular momenta up to $j_{\rm max}$=3 and total 3N angular momenta up
to $J_{\rm max}$= 15/2. It turned out to be sufficient to keep the 3N force
different from zero for total 3N angular momenta up to $J$= 7/2. The standard
 nonrelativistic form of the single nucleon current
operator~\cite{Ishikawa1994} is
supplemented by exchange currents via the Siegert approximation. We use it
in the form as detailed in~\cite{ref.15}.
In our treatment electric and magnetic multipoles are kept to a
very high order (6--7) and no long wave length approximation is used. 
Both ingredients are important as has been shown for instance  in~\cite{Trento}.
The
formalism is performed throughout in momentum space.
As is well known the Siegert approach corrects for many body currents only in the electric
multipoles. Available models for two- body currents should then be added for the magnetic
multipoles. This, however, is not yet included in this work. On the other hand we also use
explicit exchange currents of the $\pi$- and $\rho$- like type consistent to the AV18 NN force.
Again this is not yet a complete approach since further pieces in the AV18 NN force have
no counterparts in two-body currents which would be also required to fulfill the
 continuity equation.
This needs further investigations~\cite{schpriv}, though
 the expectations are
that with the $\pi$- and $\rho$- like parts the dominant currents are taken into account.
 If the
continuity equation would be fulfilled in relation to all parts of AV18 the Siegert approach
with respect to the electric multipoles would be essentially equivalent to these explicit
MEC's (except of additional 3N force effects included in the Siegert approach and less
important terms of higher multipolarities, see~\cite{ref.18}). Our aim here is not to forward the
theory of the electromagnetic current operator but to apply what is often called " the
standard model of nuclear physics" to the  complex 2-body photodisintegration
or pd capture  processes, which has not been done
before to the best of our knowledge.

\begin{figure}[h!]
\leftline{\mbox{\epsfxsize=150mm \epsffile{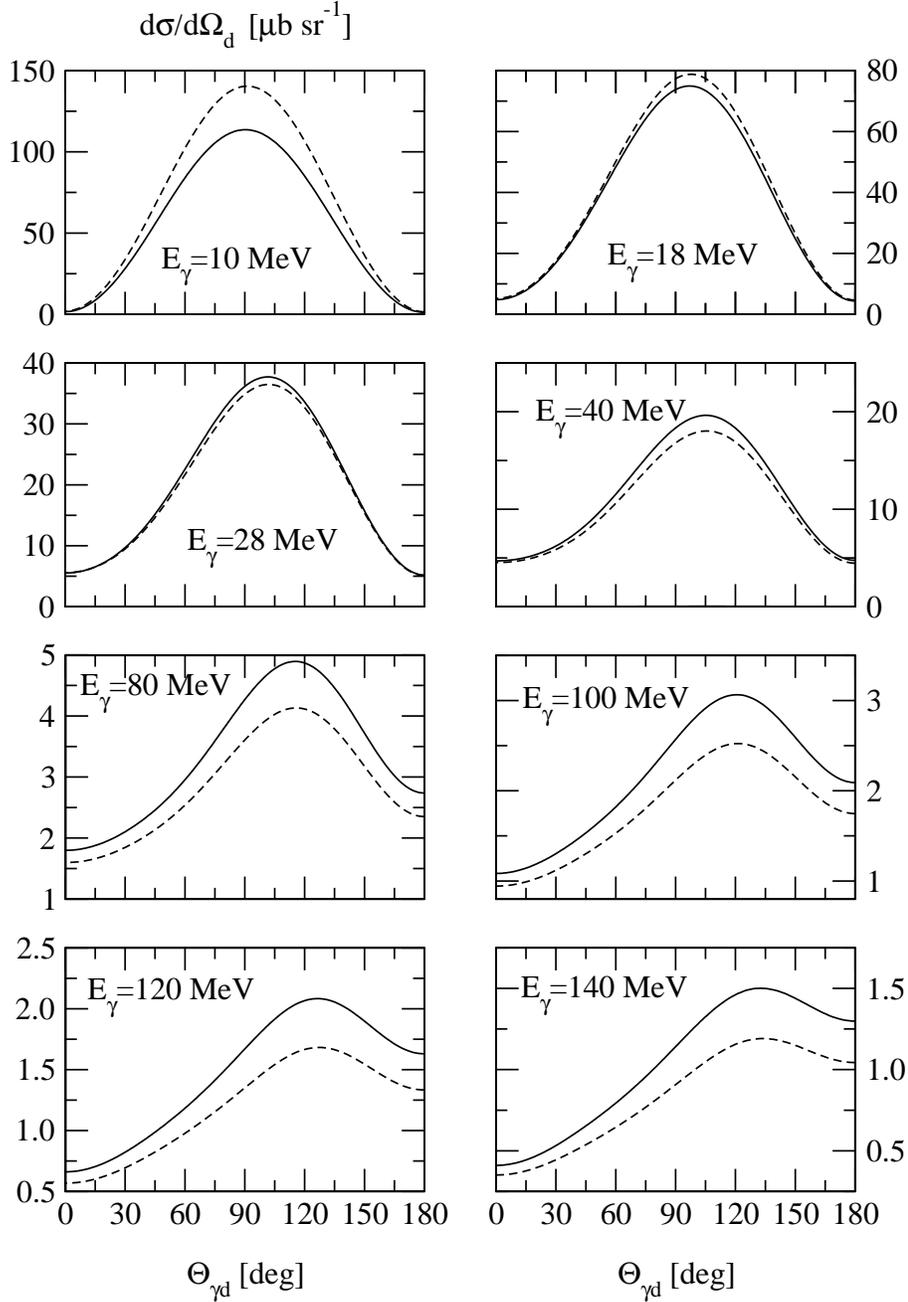}}}
\caption[ ]
{Deuteron laboratory angular distribution for the process
$^3{\rm He}(\gamma,d)p $ at different photon energies $E_\gamma$.
Curves show results of calculations with the AV18 NN
and Urbana~IX 3NF forces (solid) and with the AV18 NN force alone (dashed).
The current is treated in the Siegert approach.
}
\label{fig1}
\end{figure}

We show in Fig.\ref{fig1} the angular distribution for pd photodisintegration
of $^3$He
against the angle between the outgoing deuteron and the incoming photon
direction in the laboratory system. The photon energies $E_\gamma$ vary between
10 and 140 MeV. At the two lower energies the cross section maximum is
decreased by adding the 3NF.  A related  effect has been seen before
in~\cite{ref.Bonn}
using different NN forces. Thereby it was found, that with increasing
 binding energy the value of the maximum decreased. This
 can be considered as a scaling behavior with the 3N binding energy. It
ceases to be valid for the higher energies, where the results including
the 3N force overtake the ones without. At about $E_\gamma$= 28 MeV the
3N force effects
for the process $ ^3{\rm He} (\gamma, d)p $
in that observable vanish.
In relation to
that scaling at low energies one can ask the question whether the 3N force contributions in
the continuum are critical for that result. To that aim we performed calculations where
we switched off the 3N force in the continuum but kept it in the $^3$He bound state. Thereby the
$^3$He binding energy did not change. (Note this is not a consistent calculation and
necessarily induces spurious effects). The effect is a decrease of the cross section of
about $16\%$ at E$_\gamma$=100 MeV in relation to the difference of the results
when the 3N force is dropped
totally. At E$_\gamma$=10 MeV the presence or absence of the 3NF in the continuum had only a
very tiny effect. Nevertheless we would like to repeat that the correct binding energy
could only be achieved by adding the 3NF to the current most modern NN forces.
In~\cite{connell} energy weighted sum rules for the A=3
photodisintegration cross sections based on the electric dipole operator have been
investigated. They link the energy dependence of the cross section
 through the integrals to expectation values of
the ground state wavefunctions, which are affected by the 3N forces and consequently depend
on the binding energy. It appears worthwhile to check the assumptions and 
approximations in~\cite{connell}
from the point of view of present day forces, wavefunctions and currents.

\begin{figure}[h!]
\leftline{\mbox{\epsfxsize=150mm \epsffile{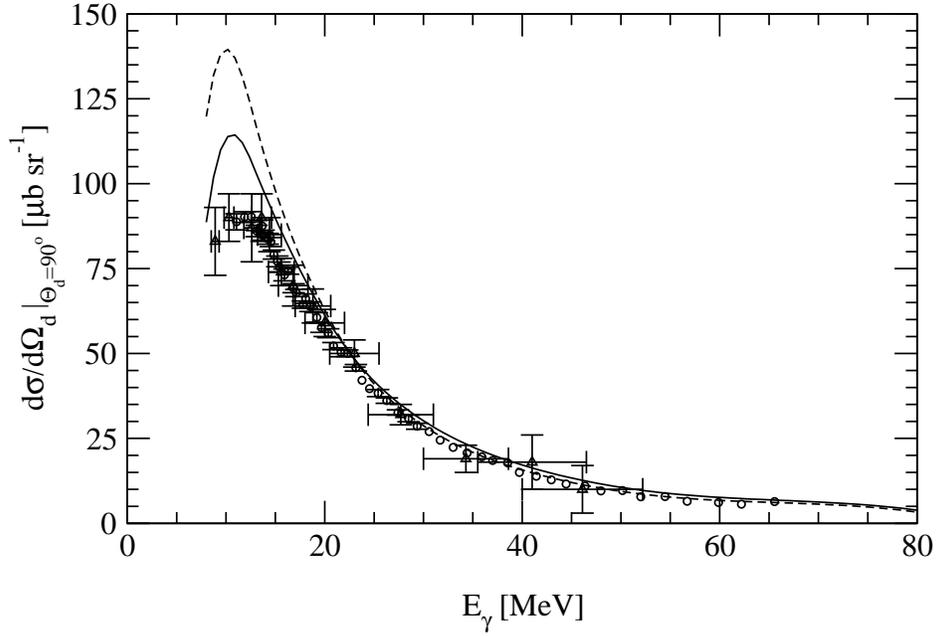}}}
\caption[ ]
{Deuteron angular distribution for the process
$^3{\rm He}(\gamma,d)p $ at the given deuteron angle
as a function of the photon energy $E_\gamma$.
Curves as in Fig.~1. Since the kinematical shift from the laboratory
to the c.m. system is not significant, we combine the data
for the 90~$^\circ$ laboratory angle (full dots with horizontal and 
vertical error bars~\cite{stewart})
with the ones for the 90~$^\circ$ c.m. angle (open circles \cite{Ticcioni}).
}
\label{fig2}
\end{figure}

Have these effects already been seen in some data? We are aware of
cross section data at the deuteron laboratory and c.m. angles
of 90~$^\circ$ as a function of $E_\gamma$\cite{stewart,Ticcioni}. They are
shown in Fig.~\ref{fig2} in comparison to our theoretical results.
We see the crossing of the theoretical curves without and with 3NF's
around 25 MeV and indeed the data support the decrease of the cross
section at lower energy values as predicted by including the 3NF.
At higher energies the effects of the 3NF appear to be somewhat too
strong in case of the Ticcioni {\em et al.} data~\cite{Ticcioni}.
It is possible
that the overshooting of the theory at the lowest energies is partially due to the neglection
of the pd Coulomb force effects in the continuum. Precise new data would be very welcome.

\begin{figure}[h!]
\leftline{\mbox{\epsfxsize=150mm \epsffile{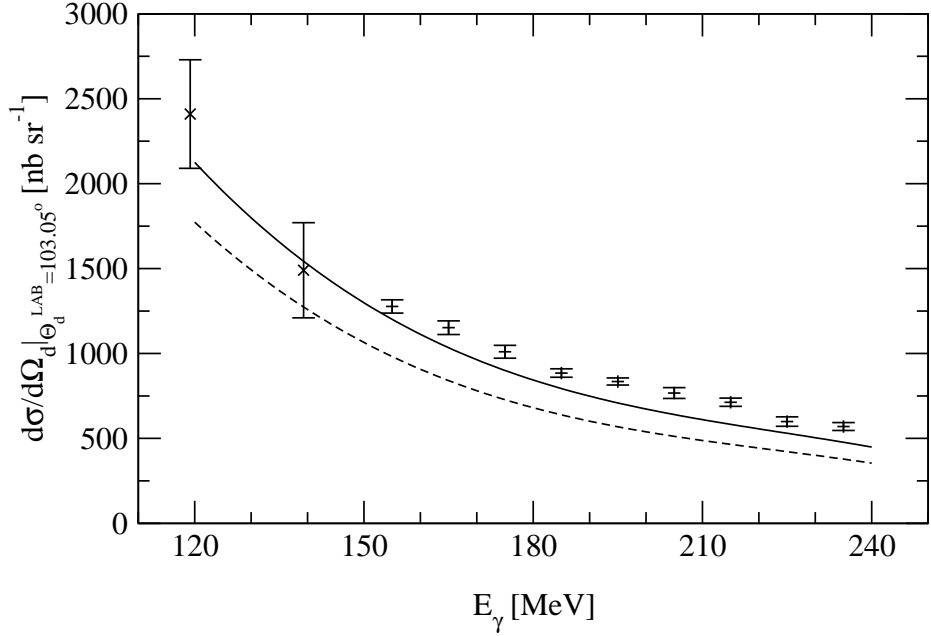}}}
\caption[ ]
{Deuteron angular distribution for the process
$^3{\rm He}(\gamma,d)p $ at given laboratory angle
as a function of the photon energy $E_\gamma$.
Curves show results of calculations with the AV18 NN
and Urbana~IX 3NF forces (solid) and with the AV18 NN force alone (dashed).
Explicit $\pi$- and $\rho$-like MEC's are included in the current operator.
Data are from \cite{sober} (x-es) and \cite{ofallon} (circles).
}
\label{fig3}
\end{figure}

For $E_\gamma$= 120 MeV and higher photon energies we are aware of
another set of data for that process~\cite{sober,ofallon}.
The deuteron laboratory angle is 103~$^\circ$ now. In this case the
addition of the 3NF is clearly supported by the data.
This is shown in Fig.~\ref{fig3}.
In this case we used the explicit $\pi$-  and $\rho$-
 like MEC's
since the energies are higher and Siegert as a low energy approximation is less suited.

\begin{figure}[h!]
\leftline{\mbox{\epsfxsize=150mm \epsffile{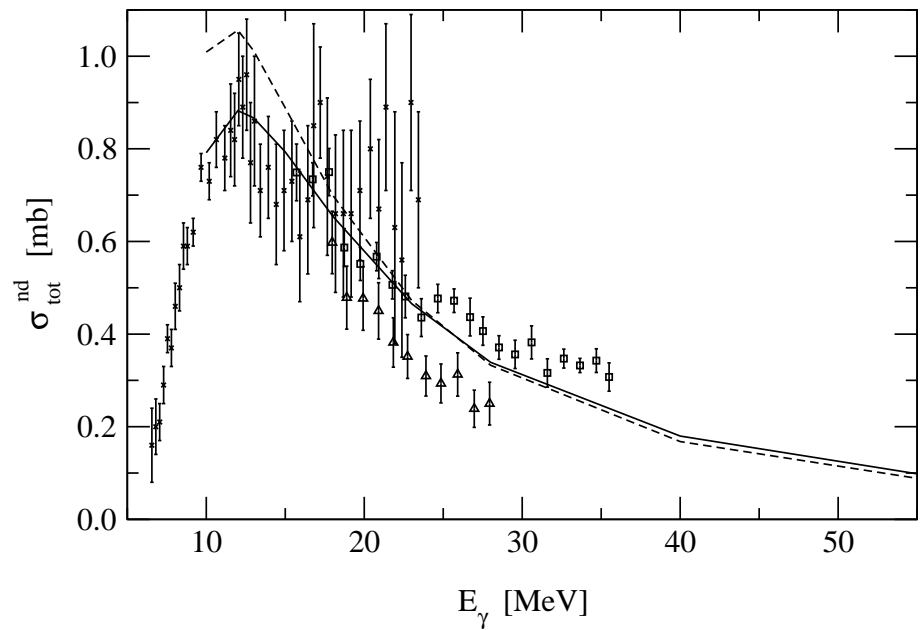}}}
\caption[ ]
{Total cross section for the process
$^3{\rm H}(\gamma,d)n $ as a function of the photon laboratory energy $E_\gamma$.
Curves as in Fig.~1.
Data are from \cite{faul} (x-es),\cite{skopik} (squares),\cite{kosiek} (triangles).
}
\label{fig4}
\end{figure}

\begin{figure}[h!]
\leftline{\mbox{\epsfxsize=150mm \epsffile{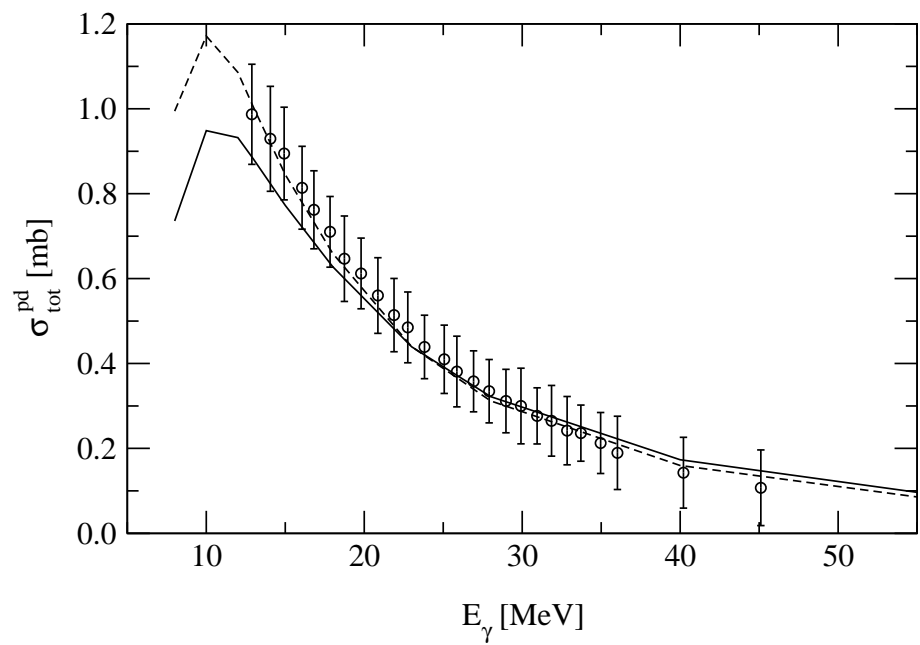}}}
\caption[ ]
{Total cross sections for the
$^3{\rm He}(\gamma,d)p$ process
as a function of the photon laboratory energy $E_\gamma$.
Curves as in Fig.~1.
Data are from \cite{kundu}.
}
\label{fig5}
\end{figure}

There are also total nd and pd cross section data for the processes
$ ^3{\rm H}(\gamma , d)n $
and
$ ^3{\rm He}(\gamma , d)p $.
We show them in Figs.~\ref{fig4} and~\ref{fig5}
based on
the Siegert approach.
Clearly the old data for the nd cross section have
too big error bars to be conclusive.
In the case of the pd cross section the inclusion of the 3NF's deteriorates
the agreement somewhat for the low photon energies.
The nd cross section data have been displayed before in~\cite{Trento}.

Summarizing, the comparison with the angular distribution data appears
to be in qualitative agreement. New improved data would be welcome and a more 
refined treatment of two-body currents is
required.

In order to provide information on the dependence of the cross sections 
on the choice of forces and currents we display in Table~I 
results for the total two-body photodisintegration cross section of $^3$He ($^3$H) 
at three energies. At 12 MeV we see 5 \% (10 \%) spreads with (without) 3NF's. 
At the higher energies the spreads are negligible which points to a
certain stability of the results and helps to identify 3N force effects. Precise data,
however, would be required.

\section{pd Capture Cross Sections}
\label{secIV}

In Ref.~\cite{ref.15} cross sections and spin
observables for pd capture have been
investigated at proton laboratory energies $E_p$ between 5 and 200~MeV
(corresponding to deuteron laboratory energies $E_d$ between 10 and 400~MeV).
The emphasis was on testing the sensitivity of pd capture
observables to changes in the choice of NN forces and to compare the
predictions of the Siegert approximation to the ones including
explicitly $\pi$- and $\rho$-like MEC's. 
Please note as pointed out above that both approaches in
the way we treat them are approximate and therefore the comparison is more of qualitative
nature. We found that at low energies Siegert
and MEC predictions are rather close together, whereas at the higher
energies differences showed up. In the context of the Siegert approach the
predictions based on different NN forces turned out to be rather close
together, which is a satisfactory result, since it demonstrates
stability. The agreement with the data was mostly good, but also clear
discrepancies were present, which call for an improvement of the
dynamical input. It is the aim of this paper to include 3NF's,
which in the previous work were only marginally investigated.

In the following we show our results for cross sections
between $E_d$= 19.8 MeV and $E_d$= 400~MeV. In all of the
following figures four theoretical curves are displayed. They are based
on the Siegert approach, the single nucleon current together with explicit
$\pi$- and $\rho$-like MEC's consistent to the NN force, the Siegert with
3NF and MEC`s with 3NF. Let us denote these four choices by a, b, a' and b',
for short.

\begin{figure}[h!]
\leftline{\mbox{\epsfxsize=150mm \epsffile{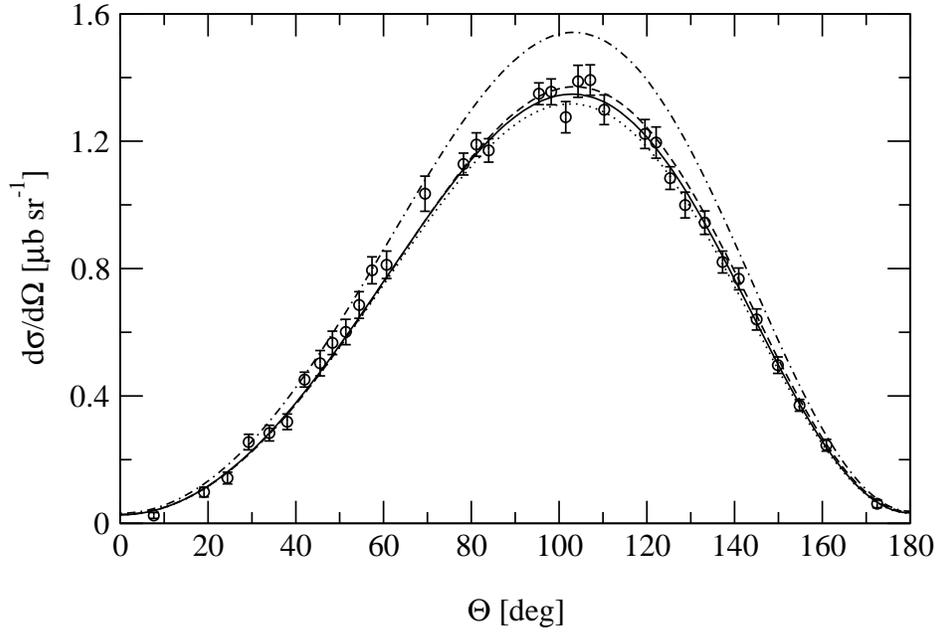}}}
\caption[ ]{
The photon angular distribution for pd capture at $E_d$= 19.8
MeV against the c.m. $\gamma$-d scattering angle.
The curves describe
the Siegert (dashed-dotted),
the single nucleon plus MEC (dashed),
Siegert with 3NF (dotted)
and the single nucleon plus MEC with 3NF (solid).
These four cases are called  a, b, a' and b'
in the text.
Data are from~\protect\cite{Belt}.
}
\label{fig6}
\end{figure}

\begin{figure}[h!]
\leftline{\mbox{\epsfxsize=150mm \epsffile{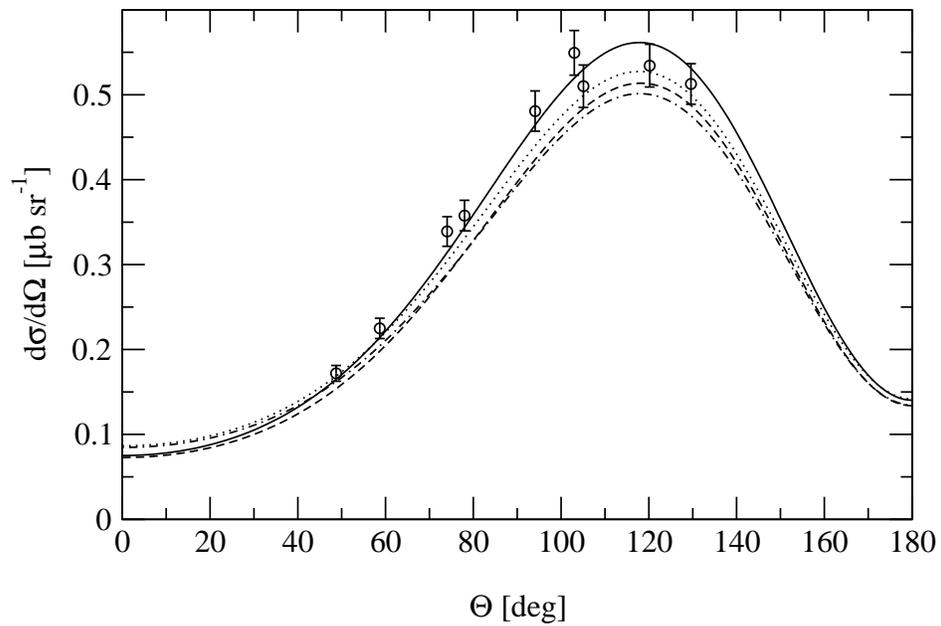}}}
\caption[ ]{
The same as in Fig.~\protect\ref{fig6} for $E_d$= 95 MeV.
Data are from~\protect\cite{Pitts}.
}
\label{fig7}
\end{figure}

We see in Fig.~\ref{fig6} the pd capture cross section
at $E_d$ = 19.8 MeV. In both cases, Siegert and explicit MEC's,
the inclusion of the 3NF
decreases the cross section; in case of Siegert the decrease is much
stronger.
 The choices a', b and b' are well within the error bars and only a is significantly too
high.
At $E_d$ = 95 MeV the cross section data are fairly well described
by all four choices. This is displayed in Fig.~\ref{fig7}.
As already seen in the Nd photodisintegration the
theoretical cross section increases by including 3NF's. This is in
agreement with our findings for pd capture~\cite{ref.15}.

\begin{figure}[h!]
\leftline{\mbox{\epsfxsize=150mm \epsffile{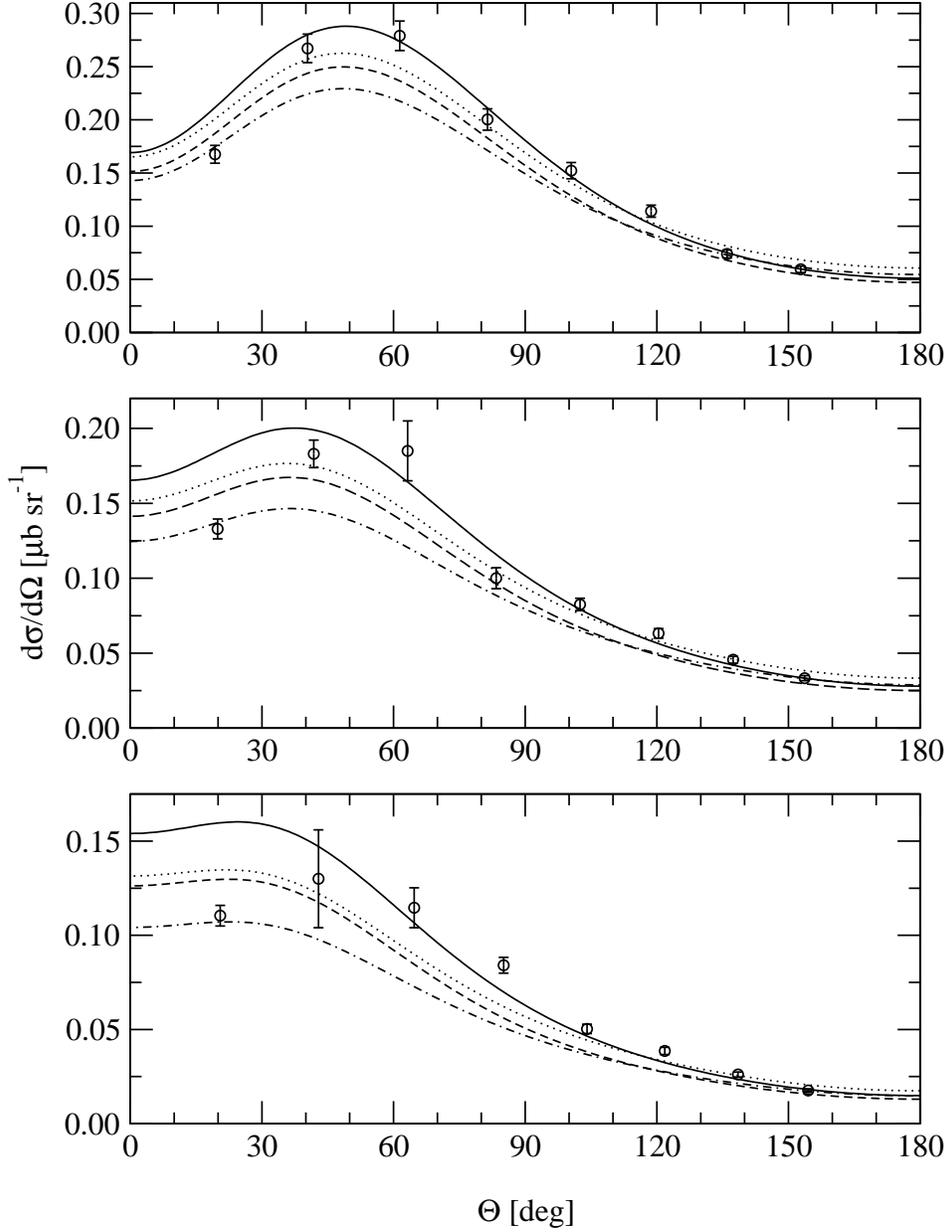}}}
\caption[ ]
{
The photon angular distribution
for pd capture
at three different
proton laboratory energies ($E_p = 100, 150$ and $200$ MeV from above to below)
against the c.m. $\gamma$-p scattering angle.
Curves as in Fig.~6.
Data are from~\protect\cite{Pickar}.
}
\label{fig8}
\end{figure}

Finally we show the cross sections
for $E_p$= 100, 150 and 200 MeV (corresponding to $E_d$= 200, 300 and 400 MeV)
in Fig.~\ref{fig8}.
The cases with the explicit MEC's and 3N forces (b') describe the data
best (except for small angles). The choice a clearly underpredicts the maxima.

\section{Summary and Conclusions}
\label{secV}

We presented the formalisms for including 3NF's into the Faddeev framework
for photodisintegration of three nucleon bound states. The resulting 
equations are solved
rigorously using high precision nuclear forces: AV18 together 
with the Urbana~IX 3NF
or CD Bonn with TM 3NF. Many-body currents are included either in the form 
of $\pi$- and $\rho$-like
exchanges related to AV18 or via the Siegert approach
 where the latter corrects only electric multipoles for
many body currents and the former does not include all two-body currents consistent to
AV18 (in the sense of fulfilling the continuity equation). Thus both ways of going beyond
the single nucleon current are approximate but currently  used in the literature.
The calculations are nonrelativistic but employ state of the art dynamics.
We posed several questions. How well do the Siegert approach
and the explicit use of the $\pi$- and $\rho$-like MEC's compare
with each other ? Our results displayed for pd capture show differences between
the two approaches which calls for improvements either by adding two-body currents for
the magnetic multipoles in the Siegert approach or by adding at least the currents beyond
the $\pi$- and $\rho$-like parts required for the consistency to AV18. Qualitatively, however,
the two approaches give similar results.

Another even more central question in that paper was to shed light
on possible 3NF effects. In case of pd photodisintegration of $^3$He
we compared theoretical predictions
without and with a 3NF. 
We found a clear signature in adding the 3NF.
The maxima are decreased at low energies and increased at high energies. The turning point is
around E$_\gamma$=28 MeV. At the low energies this can be considered as a scaling effect with
the $^3$He binding energy  but one has to note that based on present day NN forces the $^3$He
binding energy can only be achieved if a 3NF is added. These 3N force effects up to about 60
MeV are too small to be verified by the presently available data. However at the lower
energies E$_\gamma$ about 10 MeV our theoretical predictions including 3N forces are clearly
too high which might be due to the neglected Coulomb force in the continuum. At the
higher energies E$_\gamma \geq 120$MeV the effects are larger and qualitatively
supported by the data. One should, however, be aware that beyond the $\pi$-threshold
 we certainly
leave the validity of our nonrelativistic framework.
In case of the pd capture at the higher energies, E$_\gamma$=100 MeV and above, explicit
use of MEC's together with the 3N force model shows a tendency to move theory better towards
the data than without 3N forces. The failure at the smaller angles shows however that some
ingredients are missing. Overall we demonstrated that 3NF's can be incorporated into such
a complex 3N reaction process and effects are visible related to the models used. An
improved theoretical treatment of many-body currents and more precise data are needed to
achieve a clearer view towards 3N force effects.


Altogether the shifts caused by the Urbana~IX 3NF on top of the AV18 NN force
and explicit MEC's is supported by most of the data we analyzed. 
The Siegert approach is less successful. The use of other force combinations 
as exemplified in the total two-body photodisintegration cross section
does not lead to alarming variations. High quality data would be very helpful to challenge theory more strongly.

\acknowledgements

This work was supported by
the Deutsche Forschungsgemeinschaft (H.K., A.N. and J.G.),
the Polish Committee for Scientific Research under Grants No. 2P03B02818
and 2P03B05622
and by the NSF under Grant \# PHY0070858.
W.G. would like to thank the Foundation for Polish Science
for the financial support during his stay in Cracow.
R.S. thanks the Foundation for Polish Science for financial support.
The numerical calculations have been performed on the Cray T90 and T3E
of the NIC in J\"ulich, Germany.


\begin{table}
\begin{center}
\begin{tabular}{|l|c|c|c|}
\hline
& $E_{\gamma}=12$ [MeV] & $E_{\gamma}=40$ [MeV] & $E_{\gamma}=120$ [MeV]
\\
\hline
AV18-Siegert         & 1.086 (1.056) & 0.160 (0.168) & 0.016 (0.015) \\
AV18-MEC    & 0.953 (0.949) & 0.156 (0.155) & 0.017 (0.015) \\
CD Bonn2000-Siegert       & 0.997 (0.980) & 0.163 (0.169) & 0.017 (0.016) \\
\hline
AV18+UrbanaIX-Siegert & 0.932 (0.882) & 0.173 (0.180) & 0.020 (0.018) \\
AV18+UrbanaIX-MEC & 0.934 (0.915) & 0.172 (0.169) & 0.020 (0.017) \\
CDBonn2000+TM'-Siegert & 0.917 (0.889) & 0.170 (0.176) & 0.020 (0.018) \\
\hline
\end{tabular}
\caption{The total cross section (in mb) for 
two-body photodisintegration
of $^3$He ($^3$H). 
}\end{center}
\end{table}


\end{document}